\documentclass[twoside,twocolumn,9pt]{article}
\usepackage{extsizes}
\usepackage[super,sort&compress,comma]{natbib} 
\usepackage[version=3]{mhchem}
\usepackage[left=1.5cm, right=1.5cm, top=1.785cm, bottom=2.0cm]{geometry}
\usepackage{balance}
\usepackage{mathptmx}
\usepackage{sectsty}
\usepackage{graphicx} 
\usepackage{lastpage}
\usepackage[format=plain,justification=justified,singlelinecheck=false,font={stretch=1.125,small,sf},labelfont=bf,labelsep=space]{caption}
\usepackage{float}
\usepackage{fancyhdr}
\usepackage{fnpos}
\usepackage[english]{babel}
\addto{\captionsenglish}{%
	
}
\usepackage{array}
\usepackage{droidsans}
\usepackage{charter}
\usepackage[T1]{fontenc}
\usepackage[usenames,dvipsnames]{xcolor}
\usepackage{setspace}
\usepackage[compact]{titlesec}
\usepackage{hyperref}

\usepackage{epstopdf}

\definecolor{cream}{RGB}{222,217,201}

\renewcommand{\vec}{\mathbf}
\usepackage{tensor}
\usepackage{amsmath}
\usepackage{amssymb}
\usepackage{amsfonts}

\makeatletter
\g@addto@macro\@verbatim\small
\makeatother 

\newcommand{\executeiffilenewer}[3]{%
	\ifnum\pdfstrcmp{\pdffilemoddate{#1}}%
	{\pdffilemoddate{#2}}>0%
	{\immediate\write18{#3}}\fi%
}

\newcommand{\diff}{\text{d}}
\usepackage{marvosym}
\usepackage{tikz}

\begin{document}
	
	\pagestyle{fancy}
	\thispagestyle{plain}
	\fancypagestyle{plain}{
		\renewcommand{\headrulewidth}{0pt}
	}
	
	\makeFNbottom
	\makeatletter
	\renewcommand\LARGE{\@setfontsize\LARGE{15pt}{17}}
	\renewcommand\Large{\@setfontsize\Large{12pt}{14}}
	\renewcommand\large{\@setfontsize\large{10pt}{12}}
	\renewcommand\footnotesize{\@setfontsize\footnotesize{7pt}{10}}
	\makeatother
	
	\renewcommand{\thefootnote}{\fnsymbol{footnote}}
	\renewcommand\footnoterule{\vspace*{1pt}%
		\color{cream}\hrule width 3.5in height 0.4pt \color{black}\vspace*{5pt}} 
	\setcounter{secnumdepth}{5}
	
	\makeatletter 
	\renewcommand\@biblabel[1]{#1}            
	\renewcommand\@makefntext[1]%
	{\noindent\makebox[0pt][r]{\@thefnmark\,}#1}
	\makeatother 
	\renewcommand{\figurename}{\small{Fig.}~}
	\sectionfont{\sffamily\Large}
	\subsectionfont{\normalsize}
	\subsubsectionfont{\bf}
	\setstretch{1.125} 
	\setlength{\skip\footins}{0.8cm}
	\setlength{\footnotesep}{0.25cm}
	\setlength{\jot}{10pt}
	\titlespacing*{\section}{0pt}{4pt}{4pt}
	\titlespacing*{\subsection}{0pt}{15pt}{1pt}
	
	\fancyfoot{}
	\fancyfoot[RO]{\footnotesize{\sffamily{1--\pageref{LastPage} ~\textbar  \hspace{2pt}\thepage}}}
	\fancyfoot[LE]{\footnotesize{\sffamily{\thepage~\textbar ~ 1--\pageref{LastPage}}}}
	\fancyhead{}
	\renewcommand{\headrulewidth}{0pt} 
	\renewcommand{\footrulewidth}{0pt}
	\setlength{\arrayrulewidth}{1pt}
	\setlength{\columnsep}{6.5mm}
	\setlength\bibsep{1pt}
	
	\makeatletter 
	\newlength{\figrulesep} 
	\setlength{\figrulesep}{0.5\textfloatsep} 
	
	\newcommand{\topfigrule}{\vspace*{-1pt}%
		\noindent{\color{cream}\rule[-\figrulesep]{\columnwidth}{1.5pt}} }
	
	\newcommand{\botfigrule}{\vspace*{-2pt}%
		\noindent{\color{cream}\rule[\figrulesep]{\columnwidth}{1.5pt}} }
	
	\newcommand{\dblfigrule}{\vspace*{-1pt}%
		\noindent{\color{cream}\rule[-\figrulesep]{\textwidth}{1.5pt}} }
	
	\makeatother
	
	\twocolumn[
	\begin{@twocolumnfalse}
		\vspace{1em}
		\sffamily
		\begin{tabular}{m{4.5cm} p{13.5cm} }

			&
			\noindent\LARGE{\textbf{Substrate curvature governs texture orientation in thin films of smectic block copolymers}} \\
			\vspace{0.3cm} & \vspace{0.3cm} \\
			& \noindent\large{Bjarke Frost Nielsen,\textit{$^{a}$} Gaute Linga,\textit{$^{b}$} Amalie Christensen,\textit{$^{a,c}$} and Joachim Mathiesen\textit{$^{a}$}} \\
			& \vspace{0.5cm} \noindent\normalsize{Self-assembly of ordered nanometer-scale patterns is interesting in itself, but its practical value depends on the ability to predict and control pattern formation. In this paper we demonstrate theoretically and numerically that engineering of extrinsic as well as intrinsic substrate geometry may provide such a controllable ordering mechanism for block copolymers films.
			We develop an effective two-dimensional model of thin films of striped-phase diblock copolymers on general curved substrates. The model is obtained as an expansion in the film thickness and thus takes the third dimension into account, which crucially allows us to predict the preferred orientations even in the absence of intrinsic curvature. 
			We determine the minimum-energy textures on several curved surfaces and arrive at a general principle for using substrate curvature as an ordering field, namely that the stripes will tend to align along directions of maximal curvature.} \\
		\end{tabular}
		
	\end{@twocolumnfalse} \vspace{0.6cm}
	
	]
	
	\renewcommand*\rmdefault{bch}\normalfont\upshape
	\rmfamily
	\section*{}
	\vspace{-1cm}

	
	\footnotetext{\textit{$^{a}$~Niels Bohr Institute, University of Copenhagen, 2100 Copenhagen, Denmark.}}
	\footnotetext{\textit{$^{b}$~PoreLab, The Njord Centre, Department of Physics, University of Oslo, P.\ O.\ Box 1048, 0316 Oslo, Norway.}}
	\footnotetext{\textit{$^{c}$~Danmarks Nationalbank, DK-1093 Copenhagen K, Denmark.}}
	
	
	\footnotetext{\Letter~E-mail addresses: bjarkenielsen@nbi.ku.dk (BFN), gaute.linga@mn.uio.no (GL), mathies@nbi.ku.dk (JM).}
	
	
	
	\section{\label{sec:introduction} Introduction}
	Thin films of block-copolymers have received strong attention in the last two decades due to their diverse nanometer-scale self-assembly properties. Their ability to form regular hexagonal and cylindrical as well as lamellar patterns makes them promising candidates for applications in microelectronics and optics as well as nanofluidics. 
	In microelectronics and the semiconductor industry, where feature-size is of the essence, much of the appeal comes from the use of thin block copolymer films as etch masks for fabrication of ultra-small circuitry elements and memory devices. ''Bottom-up'' self-organization of block co-polymers promise to continue the miniaturization to length scales where traditional ''top-down'' lithography ceases to be feasible.\cite{Stoykovich2006,Gu2013} 
	Cylindrical phase block copolymers allow for manufacture of nanoporous membranes for ultrafiltration and molecular sieves \cite{Yang2008,Gu2015,Ahn2014,Yang2006,Xu2003} as well as superhydrophobic materials in nanofluidics.\cite{Checco2014}\\
	Lamellar and cylindrical phase block copolymer films have been demonstrated as viable templates for microelectronic circuitry and polarizing grids as well.\cite{Mansky1996,Park1997,Pelletier2006,Hong2007,ThurnAlbrecht2000}
	
	For most of these applications, a high degree of long-range order and control over macroscopic patterning is desirable. In practice, this is complicated by the formation of defects and microdomains.
	Different experimental techniques have been developed in attempts to avoid defects and obtain a macroscopic order. One such method is chemoepitaxy, where the substrate is pretreated with another chemical species, thus using the interfacial energy to facilitate the formation of long-range ordered patterns \cite{Edwards2004,Rockford1999,Cheng2008}. Shearing flow \cite{Angelescu2004,Pelletier2006,Hong2007} as well as applied electric fields \cite{Morkved1996,ThurnAlbrecht2000} have also been used with some success. Perhaps the most obvious approach to annihilating defects is annealing -- heating to near the order-disorder transition temperature and subsequently cooling. A more sophisticated version of this is the sweeping temperature gradient method, which has also proven relatively effective.\cite{Yager2010,Mita2007}
	Our work focuses on using curvature as an ordering field -- \textit{i.e.}\ using substrate topography to control the macroscopic order of lamellar patterns, analogously to an external field. Experimental studies have already shown this \textit{graphoepitaxy} technique to be a viable method to control microdomain formation. \cite{Park2009,Segalman2001,Jeong2009,Xiang2005,Cheng2002} 
	However, for this technique to be generally applicable, we must understand how to design substrates to favour the formation of specific patterns and -- conversely -- which types of pattern formation to expect as a function of substrate geometry. Our focus is on the smectic-symmetry stripe patterns obtained from compositionally symmectric diblock copolymers.
	
\begin{figure}[htb]
		\includegraphics[width=0.5\textwidth]{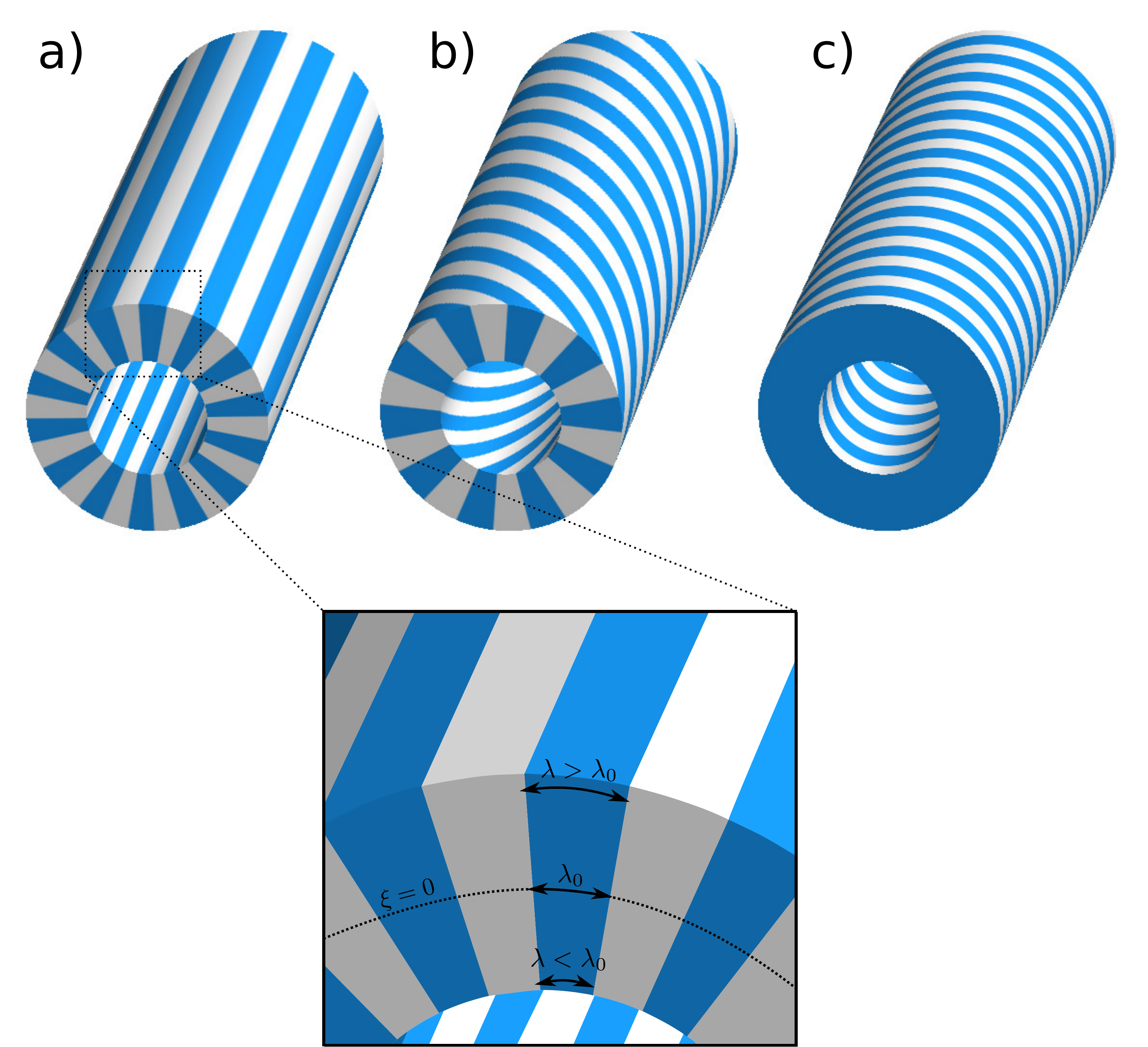}
		\caption{\label{fig:fig1} \textbf{Stripe textures on a cylindrical surface,}  with different texture
			orientations, given by the angle $\alpha$ between the stripe direction
			and the axial direction.\\
			(a) When the stripe texture runs along
			the cylinder axis $\alpha = 0$, the finite thickness of the layer and the
			curvature of the cylinder results in a slight increase of the stripe
			wavelength $\lambda$ with the radial coordinate, see inset.\\
			(b) Also the
			stripes with orientation $\alpha = \pi/4$ experience an increase of the
			wavelength in the radial direction, although the effect is smaller.\\
			(c) When the stripes run around the cylinder, $\alpha = \pi/2$ they are
			not affected by the curvature. This figure is inspired by \cite{Pezzutti2015}.}
	\end{figure}

	In this paper we consider a free energy which is dominated by the deviation of the stripe spacing from its preferred value, in accord with the approach of Pezzutti \textit{et al.}\cite{Pezzutti2015}
	Our strategy is to formulate a free energy which takes into account not only the intrinsic geometry, but also the extrinsic geometry. The latter comes into play due to the fact that the co-polymer film has a non-zero thickness. While the film is thin, the extension into the third dimension nonetheless has implications for the minimum-energy pattern on any given curved surface. Consider, as an example, the thin striped layers in Fig.\ \ref{fig:fig1}(a,b). Due to bending of the surface, the stripe spacing is forced to vary across the thickness of the film, leading to the layer being simultaneously under compression and dilation. Hence, even though the film is thin, the third dimension cannot be neglected as it effectively couples the free energy to the extrinsic curvature of the surface.
	
	By performing a systematic expansion in the thickness of the film, we obtain a two-dimensional effective theory which leads to an explicit coupling between extrinsic geometry and pattern formation.. We perform computer simulations of this system, assuming the dynamics to consist of the relaxation towards equilibrium of a conserved order parameter.
	We find that in situations where the intrinsic curvature vanishes, such as on cylinders, on ridges and in trenches, the extrinsic curvature can serve to orient the pattern in a controllable fashion. Consequently, the coupling to extrinsic geometry is indispensable in such situations.
	In cases where there is only appreciable Gaussian curvature, such as the saddle geometry, the intrinsic geometry picks out a preferred direction. In situations where both types of curvature are present we find that the extrinsic and intrinsic features may work together in orienting the pattern.
	As such, we show that there are situations where extrinsic curvature is merely a contributing factor and others where it is the crucial component in determining the macroscopic order.

	There have been previous attempts at modeling the effect of curvature on lamellar phase block co-polymer assembly, but none which correctly incorporate the effects of film thickness and coupling to extrinsic curvature for general surfaces.   
	Several authors have developed models \cite{Santangelo2007,Kamien2009}, where both intrinsic and extrinsic bending of the stripes themselves is energetically penalized.
	Intrinsic bending occurs when the stripes deviate from geodesics of the surface. Extrinsic bending, on the other hand, occurs when the stripes bend in three-dimensional space. As an example, consider the two-dimensional top layers of the cylindrical films in Fig.\ \ref{fig:fig1}. The stripes running along the cylinder (Fig.\ \ref{fig:fig1}a) have neither intrinsic nor extrinsic bending, whereas stripes running around the cylinder (Fig.\ \ref{fig:fig1}c) have no intrinsic bending but \textit{do} have extrinsic bending because they are curved in three-dimensional space. The type of model employed in \cite{Santangelo2007,Kamien2009} implies that stripes prefer to be straight in 3D and that running along the cylinder as in Fig.\ \ref{fig:fig1}a is preferred.
	
	The starting point for our expansion is a Brazovskii-type free energy. This type of free energy has been used before to model the effects of curvature on block copolymer configurations.\cite{Pezzutti2015,Matsumoto2015,Vega2013}
	Pezzutti \textit{et al.}\cite{Pezzutti2015} employ a covariantized Brazovskii surface free energy and perform a finite-thickness expansion specifically in the case of the cylinder and thus arrive at the conclusion that the preferred stripe direction is \textit{around} the cylinder, as in Figure \ref{fig:fig1}c. However they do not derive a general finite-thickness model for arbitrary surfaces and, furthermore, investigate only surfaces of vanishing Gaussian curvature. 
	Matsumoto \textit{et al.}\cite{Matsumoto2015} employ the same type of block copolymer free energy and couple it to a Canham--Helfrich membrane model of the substrate.
	By covariantizing the free energy, the metric -- and thus the intrinsic geometry -- naturally couples to the copolymer pattern. This leads to a model that predicts stripe patterns running perpendicular to substrate wrinkles for nonzero Gaussian curvature. However, since vanishing thickness is assumed, the coupling of the phase field to the \textit{extrinsic} curvature is not captured. As such, this model cannot predict a preferred orientation of stripes in e.g. a cylindrical geometry. Interestingly, Matsumoto \textit{et al.}\cite{Matsumoto2015} also allow the surface to adapt to the copolymer pattern by assuming a relaxational dynamics of the height field.
	Vega \textit{et al.}\cite{Vega2013} take a different approach, namely three-dimensional (3D) simulation of a Brazovskii-type model confined to a thin, curved patch of 3D space. They arrive at the prediction that the stripes tend to run around the cylinder. Their approach is simple in principle, requiring no covariantization or finite-thickness expansion of the free energy, but it has the disadvantage of not making the curvature-coupling explicit and of requiring simulation of a large number of degrees of freedom.
	
	It is thus clear that attempts at modeling the effects of curvature on block copolymer stripe patterns have led to contradictory results. However, experiments \cite{Vega2013,Matsumoto2015,Vu2018} may shed light on the features one should expect from a successful model of these phenomena.
	The simplest experimental paradigm in this regard is the cylinder, since it exhibits uniform extrinsic curvature while possessing no intrinsic curvature, thus allowing a separation of the effects owing to extrinsic geometry.
	In \cite{Vega2013}, polystyrene-\textit{block}-poly(ethylene-\textit{alt}-propylene) diblock copolymers were annealed on a substrate with trenches of vanishing Gaussian curvature, and it was clearly shown that the in-plane striped pattern tends to orient itself perpendicularly to the trenches. In \cite{Matsumoto2015}, the same type of block copolymer were deposited on more topographically diverse substrates, and the same tendency was seen. \\
	In \cite{Vu2018} the authors perform experiments with polystyrene-\textit{block}-poly(ethylene-\textit{alt}-propylene) diblock copolymers on both a ridge-like geometry (with vanishing Gaussian curvature) and a bumpy geometry consisting of numerous Gaussian-like smooth bumps. For the cylindrical geometry, they find that the block copolymer cylinders tend to align along the direction of curvature. For the bumpy substrate, they find that both directions of principal curvature constitute preferred orientations for the block copolymer pattern. This is in accordance with what our model predicts, as will be shown in this paper, namely that stripes preferentially align with curvature.
	
	The paper is organized as follows. In Section \ref{sec:model1}, we discuss the free energy functional which is the starting point of our description and the motivation for developing a finite-thickness model. In Section \ref{sec:model2}, we first describe the general strategy and then perform the expansion. Section \ref{sec:Results} is devoted to studying pattern formation on different geometries by numerical simulations of the model. We go on to extract the general features of the ordering mechanism and discuss the implications for pattern formation. In Section \ref{sec:conclusions} we make our closing remarks.
	
	\section{\label{sec:model1} The Free Energy}
	The Brazovskii model \cite{Brazovskii1975} and closely related Phase Field Crystal models \cite{Elder2002,Elder2004} have been applied to a broad range of systems undergoing pattern formation and selection of a specific length scale. These Brazovskii-type models have previously been employed to describe block copolymers \cite{Matsumoto2015,Pezzutti2015,Yamada2008,VillainGuillot1998,Zhang2014} but the approach is very general and also nucleation and pattern formation processes \cite{Pusztai2008,Elder2012,Elder2010}, crystal defect dynamics \cite{Elder2002,Pezzutti2011,Tarp2014,Skaugen2018}, grain boundary melting \cite{Bjerre2013,Tarp2015,Mellenthin2008} and liquid crystals \cite{Lowen2010,Wittkowski2010} have been studied. A Brazovskii-type model was also famously shown by Swift and Hohenberg\cite{swift_hydrodynamic_1977} to describe Rayleigh--Bénard convection.
	
	The Brazovskii mean field free energy $F(\psi)$ is a Ginzburg--Landau expansion in the order parameter $\psi(\vec{x})$. We will work with the corresponding free energy density $f=F/V$:
	\begin{align}
		f(\psi) = \frac{1}{V} \int \text{d}V \left[2(\nabla^2 \psi)^2 -2 \vert\nabla\psi\vert^2 + \frac{\tau}{2}\psi^2 + \frac{1}{4}\psi^4\right],\label{eq:brazovskiienergy}
	\end{align}
	where $V$ is volume, $\psi(\vec x) = \phi(\vec x)-\phi_0$ measures the local deviation from the average composition $\phi_0$ at the critical temperature $T_c$. The model has one parameter, the reduced temperature $\tau = (T_c-T)/T_c$. We assume $\phi_0=0$ throughout, since we study the compositionally symmetric lamellar phase. 
	
	The negative sign of the gradient-squared in Eq.\ \eqref{eq:brazovskiienergy} makes spatial modulations of the order parameter field $\psi$ energetically favorable. In combination with the positive Laplacian-squared, the gradient-squared favors a specific wavelength $\lambda = 2\pi \sqrt{2}$. To see this, consider the free energy density of a field $\psi = \psi_0 \sin(q_0 x)$:
	\begin{align}
		f(\psi) = (q_0^4 - q_0^2)\psi_0^2 + \frac{\tau}{2}\psi_0^2 + \frac{3}{32}\psi_0^4.
	\end{align}
	The free energy density is minimized for $q_0 = 1/\sqrt{2}$ resulting in a characteristic wavelength $\lambda = 2\pi\sqrt{2}$. Any deviation from this spacing of the stripe pattern is energetically penalized. Since Brazovskii-type models favour a single wavelength, the simulated profiles most closely resemble block copolymers in the \textit{weak segregation limit} where the composition profile (density of either component) is approximately sinusoidal \cite{Castelletto2004}, but the patterns themselves are more general.
	
	In the current work, we focus on thin films on curved surfaces. A simple way to describe the free energy of the thin film is to consider the two-dimensional surface version of Eq.\ \eqref{eq:brazovskiienergy} where all derivatives have been replaced with their covariant surface equivalents:
	\begin{align}
		f(\psi) = \frac{1}{\tilde A} \int \text{d}\tilde A \left[2(\tilde\nabla^2 \psi)^2 -2 \vert\tilde\nabla\psi\vert^2 + \frac{\tau}{2}\psi^2 + \frac{1}{4}\psi^4\right]\label{eq:brazovskiisurfaceenergy}.
	\end{align}
	Here, $\tilde\nabla$ denotes a covariant surface derivative on the surface $\tilde S$ and $\text{d}\tilde A$ is the area element of the curved surface. The strategy of replacing bulk derivatives with their surface equivalent has been applied to crystallization on curved surfaces using the related Phase Field Crystal model \cite{Backofen2010,Zhang2014} as well as in treatments of
	nematic crystals on curved surfaces using the Frank energy .\cite{Kamien2009,Nelson2002,LopezLeon2011} 
	Replacing the bulk derivatives with their surface equivalents preserves the optimal wavelength $\lambda$. This covariant formulation introduces a coupling between stripe orientation and intrinsic curvature which is geometrically clear: if the surface is intrinsically curved in some direction, the stripe pattern will effectively be stretched (or compressed) along this direction. Stretching the pattern \textit{along} the stripe direction does not affect the spacing, while stretching orthogonal to the stripes does. Thus the effect is to align the stripes along the direction of maximal intrinsic curvature.
	
	However, this approach does not take the third dimension into account and results, for example, in all stripe orientations on a cylinder being equally favorable, since the cylinder has vanishing intrinsic (Gaussian) curvature. This is not a proper description of the striped phase, which can be seen by considering Fig.\ \ref{fig:fig1}. Whereas all layers in Fig.\ \ref{fig:fig1}c have the same wavelength, this is not the case for the configuration in Fig.\ \ref{fig:fig1}a, where the wavelength increases with the radial coordinate. Thus the configuration in Fig.\ \ref{fig:fig1}c should have the lowest free energy.
	
	To properly account for the third dimension, we will start with the three-dimensional free energy density in Eq.\ \eqref{eq:brazovskiienergy} and expand it in the thickness of the film, to obtain a two-dimensional free energy density which takes both the intrinsic and extrinsic curvature of the surface into account.
	
	\section{\label{sec:model2} Coupling between substrate curvature and texture orientation}
	\subsection{\label{sec:geomsetup}Geometrical setup} 
\begin{figure} 
		\centering
		\def\svgwidth{\columnwidth}
		\begingroup%
		\makeatletter%
		\providecommand\color[2][]{%
			\errmessage{(Inkscape) Color is used for the text in Inkscape, but the package 'color.sty' is not loaded}%
			\renewcommand\color[2][]{}%
		}%
		\providecommand\transparent[1]{%
			\errmessage{(Inkscape) Transparency is used (non-zero) for the text in Inkscape, but the package 'transparent.sty' is not loaded}%
			\renewcommand\transparent[1]{}%
		}%
		\providecommand\rotatebox[2]{#2}%
		\newcommand*\fsize{\dimexpr\f@size pt\relax}%
		\newcommand*\lineheight[1]{\fontsize{\fsize}{#1\fsize}\selectfont}%
		\ifx\svgwidth\undefined%
		\setlength{\unitlength}{331.02171678bp}%
		\ifx\svgscale\undefined%
		\relax%
		\else%
		\setlength{\unitlength}{\unitlength * \real{\svgscale}}%
		\fi%
		\else%
		\setlength{\unitlength}{\svgwidth}%
		\fi%
		\global\let\svgwidth\undefined%
		\global\let\svgscale\undefined%
		\makeatother%
		\begin{picture}(1,0.44224414)%
		\lineheight{1}%
		\setlength\tabcolsep{0pt}%
		\put(0,0){\includegraphics[width=\unitlength,page=1]{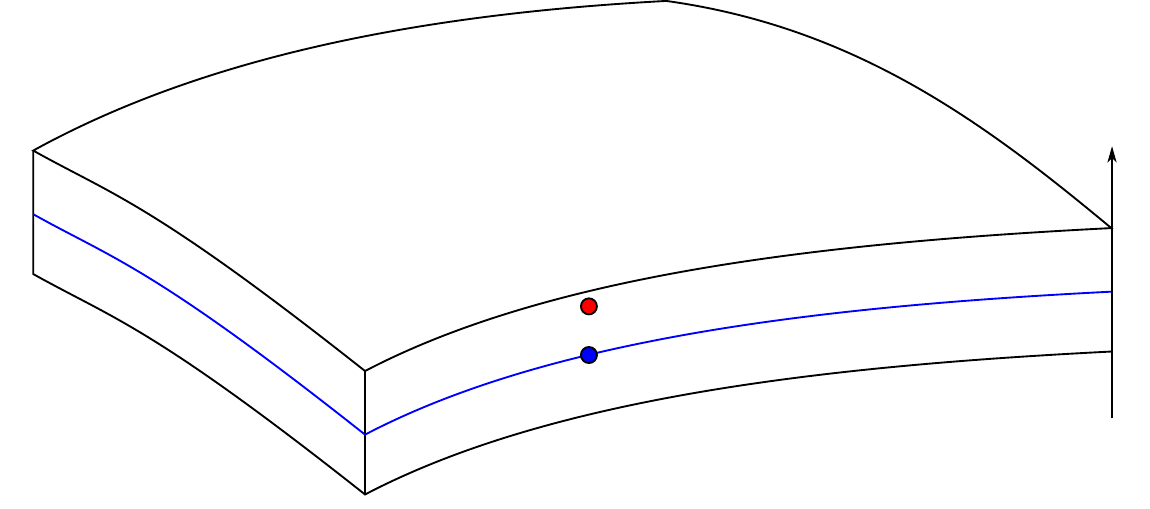}}%
		\put(0,0){\includegraphics[width=\unitlength,page=2]{parallelsurfaces_vector.pdf}}%
		\put(0,0){\includegraphics[width=\unitlength,page=3]{parallelsurfaces_vector.pdf}}%
		\put(0.43145945,0.00854149){\color[rgb]{0,0,0}\makebox(0,0)[lt]{\lineheight{1.25}\smash{$\textcolor{blue}{\tilde{\mathbf{p}}(u,w)}$}}}%
		\put(0.978 ,0.32){\color[rgb]{0,0,0}\makebox(0,0)[lt]{\lineheight{1.25}\smash{\large $\xi$}}}%
		\put(0.98311526,0.23863372){\color[rgb]{0,0,0}\makebox(0,0)[lt]{\lineheight{1.25}\smash{$\frac{h}{2}$}}}%
		\put(0.98325875,0.175){\color[rgb]{0,0,0}\makebox(0,0)[lt]{\lineheight{1.25}\smash{$0$}}}%
		\put(0.98303013,0.125){\color[rgb]{0,0,0}\makebox(0,0)[lt]{\lineheight{1.25}\smash{$-\frac{h}{2}$}}}%
		\put(0.52447471,0.14508616){\color[rgb]{0,0,0}\makebox(0,0)[lt]{\lineheight{1.25}\smash{$\tilde{\mathbf{n}}(u,w)$}}}%
		\put(0.31326399,0.26456767){\color[rgb]{0,0,0}\makebox(0,0)[lt]{\lineheight{1.25}\smash{$\textcolor{red}{\mathbf{p}(u,w,\xi)} = \textcolor{blue}{\tilde{\mathbf{p}}(u,w)} + \xi \tilde{\mathbf n}(u,w)$}}}%
		\put(-0.007, 0.3){\color[rgb]{0,0,0}\makebox(0,0)[lt]{\lineheight{1.25}\smash{\large $\textcolor{red}{S}$}}}%
		\put(-0.007, 0.23){\color[rgb]{0,0,0}\makebox(0,0)[lt]{\lineheight{1.25}\smash{\large $\textcolor{blue}{\tilde S}$}}}%
		\put(0.07674343,0.37226746){\color[rgb]{0,0,0}\makebox(0,0)[lt]{\lineheight{1.25}\smash{\LARGE $\Omega$}}}%
		\end{picture}%
		\endgroup%
		\caption{\textbf{The geometric setup for the expansion}. $\Omega$ is the entire three-dimensional volume of the film while $\tilde S$ defines the midsurface (blue) and $S$ represents a surface (red) within $\Omega$, separated from $\tilde S$ by a distance $\xi$ along the normal vector $\tilde{\mathbf{n}}$.}
		\label{fig:geometricsetup} 
	\end{figure}

We consider a thin three-dimensional region $\Omega$ of thickness $h$ around a regular compact surface $\tilde S$ -- see Figure \ref{fig:geometricsetup}. We define $\tilde{\vec{n}}$ to be the unit normal vector field to the surface $\tilde S$. The volume $\Omega$ is described by the three-dimensional position vector $\vec{p}(u, w,\xi)$ parametrized by the three internal parameters $(u, w, \xi)$:
	\begin{align}
		\vec{p}(u, w,\xi) = \tilde{\vec{p}}(u, w) + \xi \tilde{\vec{n}}(u, w), \label{eq:pointexpansion}
	\end{align}
	where $\tilde{\vec{p}}$ is the normal projection of the point $\vec p$ onto $\tilde S$. The distance between $\vec{p}$ and the surface $\tilde S$ along the normal $\tilde{\vec{n}}$ at a point $\tilde{\vec{p}}$ is given by $\vert\xi\vert$. The surface is of thickness $h$ and thus $\xi \in [-h/2;h/2]$.
	The tangent vectors at the point $\tilde{\vec{p}}(u,w) \in \tilde S$ are
	\begin{align}
		\tilde{\vec{a}}_i = \partial_i\tilde{\vec{p}},
	\end{align}
	where the tilde indicates that the tangent vectors belong to the surface $\tilde S$ and the index $i$ runs over the reference coordinates $u$ and $w$.
	
	The induced metric (first fundamental form) on the surface $\tilde S$ is
	\begin{align}
		\tilde g_{ij} &= \tilde{\vec{a}}_i \cdot \tilde{\vec{a}}_j,
	\end{align}
	where $\cdot$ indicates the standard Euclidean inner product in $\mathbb{R}^3$.
	The metric determinant will be denoted $\tilde g$. The metric inverse is $\tilde g^{ij}$ and defined such that $\tilde g^{ik} \tilde g_{kj} = \delta^i_j$ where repeated indices indicate summation (Einstein convention).
	The metric and its inverse can be used to raise and lower indices. The curvature tensor\footnote{Not to be confused with either the Riemann or Ricci curvature tensors which are purely intrinsic.} (second fundamental form) of the surface $\tilde S$ is:
	\begin{align}
		K_{ij} &= \tilde{\vec{n}} \cdot \partial_i \tilde{\vec{a}}_j. \label{eq:extrinsiccurvaturetensor}
	\end{align}
	We denote the two principal curvatures at a point $\tilde{\vec{p}}$ as $\kappa_1(\tilde{\vec{p}})$ and $\kappa_2(\tilde{\vec{p}})$ respectively. If we define the local curvature length scale, $l(\tilde{\vec{p}})$ and the global curvature length scale $\ell$ as:
	\begin{align}
		l(\tilde{\vec{p}}) = \text{min}\left[ \frac{1}{\kappa_1(\tilde{\vec{p}})}, \frac{1}{\kappa_2(\tilde{\vec{p}})} \right], ~~~~ \ell = \underset{{\tilde{\vec{p}} \in \tilde S}}{\text{min}} \, l(\tilde{\vec{p}}),
	\end{align}
	then the requirement of the volume $V$ being a thin shell can be formulated as 
	\begin{align}
		\left(\frac{h}{\ell}\right)^2 \ll 1.
	\end{align}
	We consider a scalar order parameter $\psi$ which is constant throughout the thickness of the shell -- something which will be important once we derive the effective two-dimensional free energy:
	\begin{align*}
		\psi(\tilde{\vec{p}} + \xi \tilde{\vec{n}}) = \psi(\tilde{\vec{p}}) ~~~ \text{for all } \tilde{\vec{p}} \in \tilde S, ~ \xi \in [-h/2;h/2].
	\end{align*}
	The goal is to express the three-dimensional free energy density described by eqn \eqref{eq:brazovskiienergy} in terms of the curvature tensor in eq \eqref{eq:extrinsiccurvaturetensor}, invariants of the surface $\tilde S$ such as the mean curvature $H$ and the Gaussian curvature $K$ and surface covariant derivatives of the fields.
	We write it as an expansion in the surface normal coordinate $\xi$. Once this has been done, the surface height $\xi \in [-h/2;h/2]$ can be integrated out to arrive at an effective two-dimensional free energy density to lowest order in the surface thickness to curvature ratio $h/\ell$.
	
	\subsection{\label{sec:theexpansion} Expansion of the free energy density}
	Geometrically, the curvature tensor $\tensor{K}{^i_j}$ expresses the rate of change of the normal vector projected onto the surface:
	\begin{align}
		\tensor{K}{_{ij}} &= \tilde{\vec{n}} \cdot \partial_i \tilde{\vec{a}}_j = - \tilde{\vec{a}}_j \cdot \partial_i \tilde{\vec{n}}
	\end{align}
	In our analysis we will assume the curvature to vary slowly compared to the characteristic wavelength such that gradients of the curvature tensor can be neglected.\\
	When combining equation \eqref{eq:pointexpansion} with $\vec{a}_i = \partial_i \vec{p}$ it is clear that the tangent vectors of the surface $S$ will involve the extrinsic curvature. The metric tensor $g_{ij} = \vec a_i \cdot \vec a_j$ then inherits this dependence on curvature:
	\begin{align}
		g_{ij} &= \tilde{g}_{ij} - 2 \xi K_{ij} + \xi^2 \tensor{K}{^k_i} \tensor{K}{_{kj}}.
	\end{align}
	This expression is exact. To second order in the small quantity $\xi/\ell$ the \textit{inverse} metric $g^{ij}$ takes the following form:
	\begin{align}
		g^{ij} &= (1-3 \xi^2 K) \tilde g^{ij} + 2 (\xi + 3 \xi^2 H) K^{ij} + \mathcal{O}(\xi/\ell)^3.
	\end{align}
	The curvature will then enter into the $\vert\nabla\psi\vert^2$ and $(\nabla^2\psi)^2$ terms of the free energy through the metric. However, the volume element itself is also affected. The invariant volume element in differential geometry is $\sqrt{g} \ \text{d}^d x$ where $g$ is the determinant of the metric.
	We imagine the volume $\Omega$ to be foliated by a series of surfaces $S$, each being a level set of $\xi$ described by \eqref{eq:pointexpansion} such that $\xi = 0$ corresponds to the midsurface $\tilde{S}$. The volume of $\Omega$ is denoted by $V$ while the surface areas of S and $\tilde S$ are denoted by $A$ and $\tilde A$, respectively. The volume element $\diff V = \diff\xi\diff A$ depends on $\xi$ through the area element $\diff A$. This dependence follows from the expansion of the metric determinant in terms of $\xi$, which is given by \cite{Deserno2004}:
	\begin{align}
		\sqrt{g} &= J_\xi \sqrt{\tilde g}, ~~~~ J_\xi = 1 - 2 H \xi + K \xi^2.
	\end{align}
	The area element of $S$ is then $\diff A = J_\xi \diff\tilde{A}$. It follows that the total volume $V$ which enters the free energy density is given by
	\begin{align*}
		V &= \int_\Omega \diff V = \int_{-h/2}^{h/2} \diff \xi \int_{\tilde{S}} \diff\tilde{A} J_\xi = \tilde{A} h + \frac{h^3}{12}\chi,
	\end{align*} 
	where $\chi \equiv \int_{\tilde S} \diff\tilde{A} K$ is the integrated Gaussian curvature which, by the Gauss--Bonnet theorem, equals $2\pi$ times the Euler characteristic for a closed surface. 
	
	The gradient-squared term is expanded as
	\begin{align*}
		J_\xi \vert \nabla\psi\vert = J_\xi g^{ij} \nabla_i\psi\nabla_j\psi = \vert\tilde\nabla\psi\vert^2 + c_1 \xi + c_2 \xi^2 + c_3 \xi^3 + \mathcal{O}(\xi/\ell)^4
	\end{align*}
	where
	\begin{align*}
		c_2 &= 2 H K^{ij} \tilde\nabla_i\psi \tilde\nabla_j\psi - 2 K \left\vert \tilde\nabla\psi \right\vert^2.
	\end{align*}
	The odd terms ($c_1$ and $c_3$) will not contribute to the effective 2D theory, since they integrate to zero over $\xi \in [-h/2,h/2]$.\\
	The Laplacian-squared term can be expanded similarly to yield
	\begin{align*}
		J_\xi \vert \nabla\psi\vert &= (\tilde\nabla^2\psi)^2 + d_1 \xi + d_2 \xi^2 + d_3 \xi^3 + \mathcal{O}(\xi/\ell)^4
	\end{align*}
	where the relevant curvature coupling term is given by
	\begin{align*}
		d_2 &= 4 (K^{ij} \tilde\nabla_i\psi \tilde\nabla_j\psi)^2 + 4 H (K^{ij} \tilde\nabla_i\psi \tilde\nabla_j\psi) \tilde\nabla^2\psi - 5K (\tilde\nabla^2\psi)^2.
	\end{align*}
	The last ingredient in the expansion is the local, polynomial part of the free energy. This depends on $\xi$ only due to the metric determinant as it appears in the volume element.

	The final effective two-dimensional energy up to and including order $(h/\ell)^3$ takes the form
	\begin{align}
		f 
		&= \frac{h\tilde A}{V} \left\{ f_{\tilde S} +k_{\tilde S} \right\}
		\label{eq:f_final}
	\end{align}
    Here, $f_{\tilde S}$ is the covariantized Brazovskii surface free energy as given by Eq.\ \eqref{eq:brazovskiisurfaceenergy}, to which this effective energy reduces when $h \to 0$. The correction $k_{\tilde S}$ due to a finite film thickness is given by:
    \begin{align*}
        k_{\tilde S} &= \frac{h^2}{12 \tilde A} \int_{\tilde S} \text{d}\tilde{A} \Big[ 8 ((K^{ij} \tilde\nabla_i \tilde\nabla_j \psi)^2 + H (K^{ij} \tilde\nabla_i \tilde\nabla_j \psi) (\tilde\nabla^2\psi))\nonumber \\
		&- 10 K(\tilde\nabla^2\psi)^2 - 4(H K^{ij}  (\tilde\nabla_i \psi) (\tilde\nabla_j \psi) - K \vert \nabla\psi\vert^2) + K (\tfrac{\tau}{2}\psi^2 + \tfrac{1}{4}\psi^4) \Big] .
    \end{align*}

    \subsection{Relaxation towards equilibrium}
    Assuming a conserved order parameter field, the relaxation in time $t$ towards equilibrium can be described by the equation
    \begin{align}
        \frac{\partial \psi}{\partial t} = M \tilde\nabla^2 \frac{\delta f}{\delta \psi},
        \label{eq:dynamics}
    \end{align}
    where $M$ is a diffusion coefficient which sets the time scale of the dynamics, and $\delta f /\delta \psi$ denotes the functional derivative of the free energy in \eqref{eq:f_final}.
    
    To study the effects of curvature on non-trivial surfaces, we solve the dynamic equation \eqref{eq:dynamics} numerically using a finite element method in space and an implicit finite difference scheme in time.
    The numerical method is described in detail in appendix \ref{sec:numerics}.
    
    Note that the equation of motion \eqref{eq:dynamics} guarantees that the free energy $f$ decreases in time, and thus the system will eventually reach at least a \emph{local} free energy minimum. 
    However, when the initial state $\psi ( \tilde{\vec{p}}, t=0)$ is sufficiently disorganized, this local minimum state may be far from the \emph{global} minimum in the free energy, which we typically seek.
    To get closer to this state, a cyclic annealing procedure, as outlined by Zhang \textit{et al.}\cite{Zhang2014}, was implemented.
    This procedure is described in more detail in Sec.\ \ref{sec:cylinder}.
	
	\section{\label{sec:Results} The effects of curvature as an ordering field}
	In order to understand how to design substrates in order to obtain specific textures, it is necessary to understand how curvature acts as an ordering field for the striped phase in specific geometries. In this section we will study the low-energy texture configurations on qualitatively different curved surfaces which exemplify the distinct configurations of Gaussian curvature $K$ and mean curvature $H$. The surfaces considered are the cylinder ($K=0$, $H \neq 0$), the saddle geometry ($K \neq 0$, $H=0$) and the Gaussian bump ($H \neq 0$, $K \neq 0$).
	
\begin{figure}[htb]
		\includegraphics[width=\columnwidth]{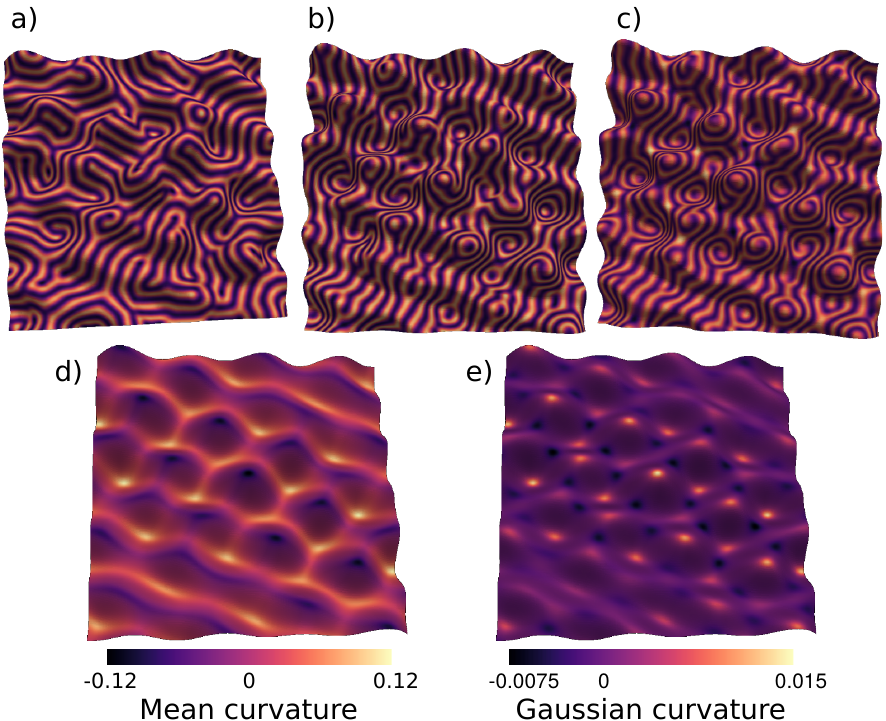}
		\caption{\label{fig:bumpy} \textbf{Effect of thickness on pattern formation.} Shown above are the results of quenching simulations on a random landscape for \textbf{(a)} $(h/\ell)^2=0$, \textbf{(b)} $(h/\ell)^2=0.49$ and \textbf{(c)} $(h/\ell)^2=1.25$ respectively. Note that the rightmost plot corresponds to $h/\ell > 1$ and thus lies outside the thin shell regime.
		\textbf{\textbf{(d)}} Mean curvature. \textbf{(e)} Gaussian curvature.}
	\end{figure}

	Before delving into the details of the mechanism on model geometries, we have simulated the model on a random landscape of sinusoidal bumps and ridges, see Figure \ref{fig:bumpy}. As film thickness increases from left to right in the figure, it becomes clear that there is a tendency for the stripes to run perpendicularly to the ridge-like features and to encircle the bumps.
	
	With this intuition in mind, let us turn to the study of pattern formation on simpler model surfaces. We take the surface to be locally parametrized in terms of coordinates $x^1=u$ and $x^2=w$ and that the coordinate curves are chosen as lines of curvature, rendering the metric as well as the curvature tensor diagonal. The curvature tensor may then be completely specified by the mean curvature $H$ and Gaussian curvature $K$:
	\begin{align}
		\tensor{K}{^i_j} &= \begin{bmatrix} H \pm \sqrt{H^2 - K} & 0 \\ 0 & H \mp \sqrt{H^2 - K} \end{bmatrix},
	\end{align}
	In order to see the role of intrinsic and extrinsic geometry separately in the finite-thickness energy contribution, it is instructive study the model in the extremal cases of vanishing Gaussian curvature ($K=0$, $H \neq 0$) and vanishing mean curvature ($K \neq 0$, $H=0$), respectively. Below we study two simple examples of such extremal geometries and solve for the preferred pattern orientation, namely the cylinder and the saddle geometry.

	We can parametrize a one-mode stripe pattern on a surface (such as those of Fig.\ \ref{fig:fig1}) as
	\begin{align}
		\psi = \psi_0 \cos[k_0 (\cos(\alpha) s_u + \sin(\alpha) s_w) ] \label{eq:modewithcurvature}
	\end{align} where $s_u$ and $s_w$ are the arc lengths along the coordinate curves of $u$ and $w$ on the surface. In the next section we will use this expression to derive preferred orientations on different geometries. 
	
	\begin{figure}[htb]
		\includegraphics{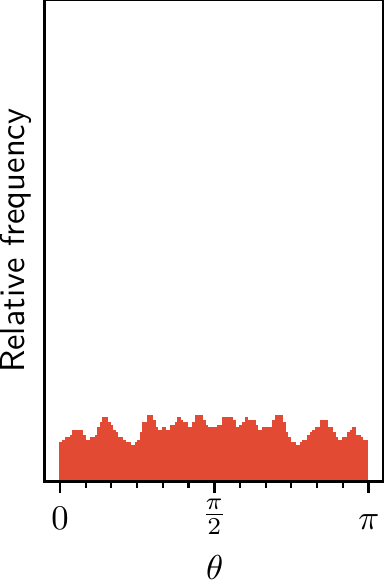}
		\includegraphics{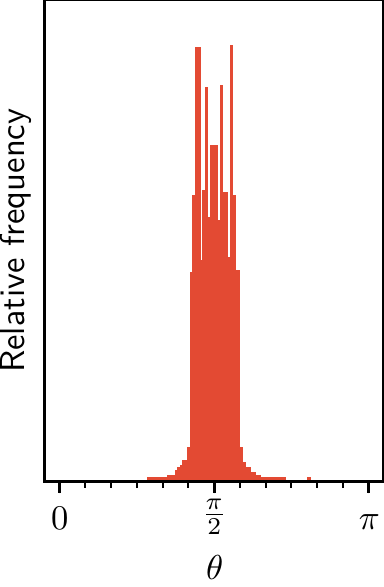}
		\caption{\label{fig:cylhist} \textbf{Average angle distribution on the cylinder after annealing. Left: $(h/\ell)^2=0$. Right: $(h/\ell)^2=0.5$}.\\
		Here $\theta=\angle(\hat{\boldsymbol{\phi}},\tilde{\nabla}\psi)$. 
		For $h>0$ there is a clear tendency for $\tilde{\nabla}\psi$ to be oriented in the axial direction (along $\hat{\vec z}$), meaning that the stripes tend to run around the cylinder (along $\hat{\boldsymbol{\phi}}$). The histograms are based on 5 (left) and 17 (right) simulations.}
	\end{figure}
	
	\subsection{\label{sec:cylinder} Cylinder}
	The cylinder is an example of a geometry satisfying $K=0$, $H \neq 0$ as described above. In this case the curvature effects are entirely extrinsic in nature, meaning that the finite-thickness contribution to the energy is crucial in breaking the symmetry between all the possible stripe orientations.
	
	We parametrize a cylindrical surface $\tilde S$ of radius $R$ and length $L$ by the cylindrical coordinates $\theta \in [0,2\pi]$, $z \in [0,L]$:
	\begin{align}
		\tilde{\vec{p}}(\theta,z) &= \begin{bmatrix} R\cos(\theta) & R\sin(\theta) & z
		\end{bmatrix}^T
	\end{align}
	Relevant geometrical quantities associated for this specific parametrization are:
	\begin{align*}
		\tilde{g}_{ij} &= \begin{bmatrix} \tilde{g}_{\theta\theta} & \tilde{g}_{\theta z} \\ \tilde{g}_{z \theta} & \tilde{g}_{zz} \end{bmatrix} = \begin{bmatrix} R^2 & 0 \\ 0 & 1 \end{bmatrix},\\
		K_{ij} &= \begin{bmatrix} R & 0 \\ 0 & 0 \end{bmatrix},\\
		K &= 0, ~~ H=1/(2R).
	\end{align*}
	As in \cite{Pezzutti2015}, we consider a striped texture making an angle $\alpha$ with the axis of the cylinder, see Fig.\ \ref{fig:fig1}. By applying Eq.\ \eqref{eq:modewithcurvature}, we obtain the following expression:
	\begin{align*}
		\psi(\theta,z) &= \psi_0 \cos[q_0(R\theta\cos(\alpha) + z \sin(\alpha))].
	\end{align*}
	Since the Gaussian curvature vanishes everywhere, the curvature contribution to the free energy in Eq.\ \eqref{eq:f_final} reduces to:
	\begin{align*}
		k_{\tilde S} &= \frac{1}{12} \left(\frac{h}{R}\right)^2 \psi_0^2 \cos^4(\alpha).
	\end{align*}
	where we have inserted the preferred wavenumber $q_0 = 1/\sqrt{2}$ in the last step. The curvature contribution to the energy is minimized when $\alpha = \pi/2$ and the stripes on every parallel surface are able to maintain the preferred lattice spacing $q_0$, as shown in Figure \ref{fig:fig1}c.
	This specific result for the cylinder has previously been derived \cite{Pezzutti2015} and is in agreement with the observation of stripe textures running perpendicular to substrate ridges.\cite{Vega2013,Matsumoto2015}
	
	We have simulated the model on a cylinder and measured the angle of the gradient $\tilde\nabla\psi$, which is of course perpendicular to the stripes. The angle histograms for the $h=0$ (vanishing thickness) and $h>0$ cases are shown in Figure \ref{fig:cylhist}. The tendency for stripes to run \textit{around} the cylinder is very clear.

	To obtain such a clear result, it was necessary to perform a cyclical heat treatment -- \textit{annealing} -- in order to decrease the number of dislocations and reach a low-energy state. Our annealing protocol consists of cycling sinusoidally between a low temperature ($\tau = 0.1$) and a high temperature ($\tau=0.99$) which lies very close to the order-disorder transition point at $\tau=1$.
	
	\subsection{\label{sec:saddle} Saddle geometry}
	\begin{figure}[htb]
		\centering
		\includegraphics[width=0.8\columnwidth]{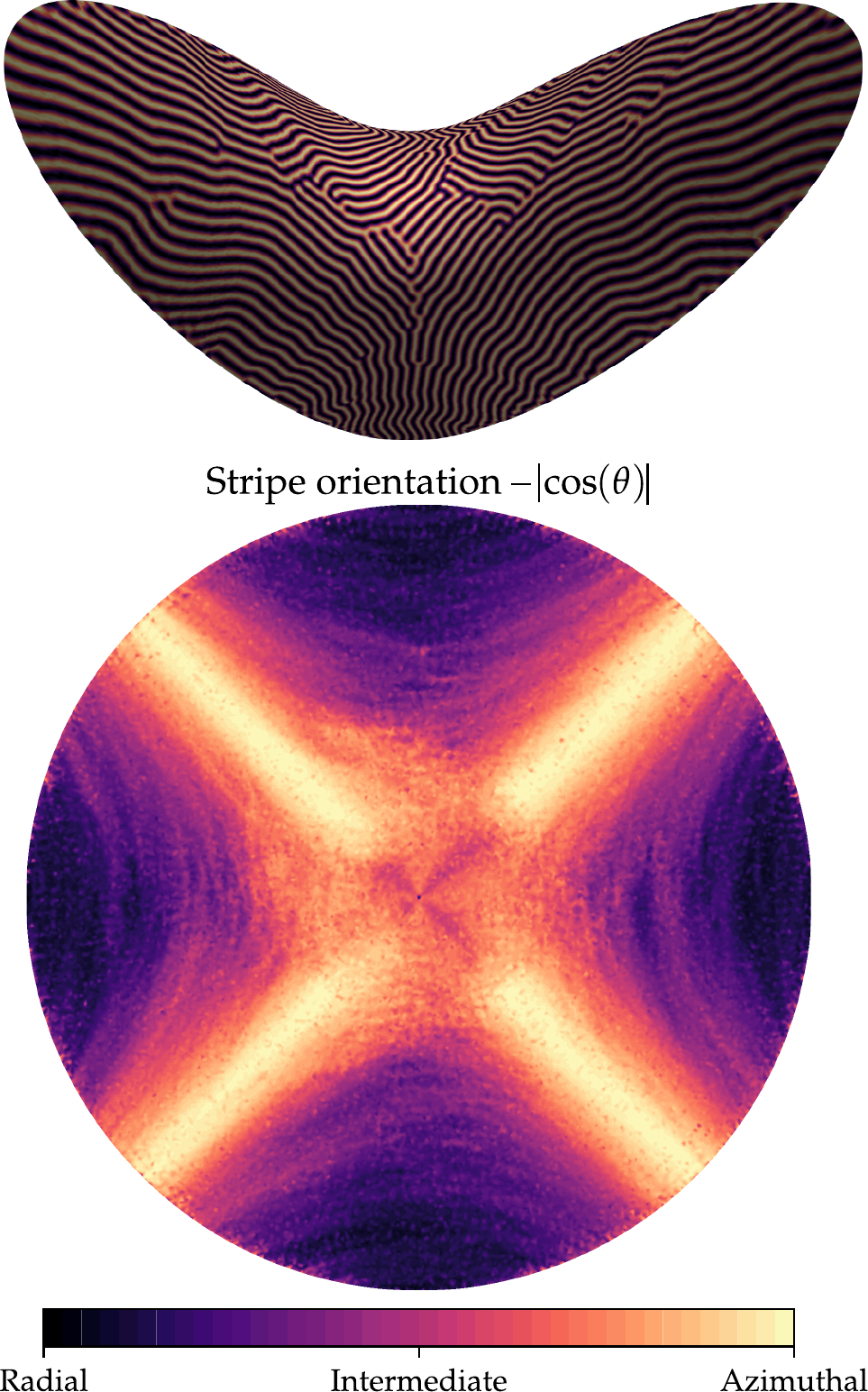}
		\caption{\label{fig:saddle} \textbf{Stripe textures in the vicinity of a saddle point.}\\
			\textit{Top:} A tilted view of a representative stripe pattern on the saddle geometry, reached by annealing.\\
			\textit{Bottom:} Spatial angle distribution, averaged over $12$ runs such as the one shown in the top figure. The color denotes $\left\vert\cos(\theta)\right\vert$ with $\theta = \angle(\hat{\vec{r}},\tilde{\nabla}\psi)$
			being the angle between the radial vector and the gradient of $\psi$ along the curved surface. These simulations were run with $(h/\ell)^2 = 0.23$.}
	\end{figure} An example of the $H = 0$, $K \neq 0$ situation can be realized in a simple saddle geometry.
	This geometry can be parametrized as $\tilde{\vec{p}}(x,y) = [x, y, \tfrac{a}{2}(y^2-x^2)]^T$ in Monge gauge. In this case the metric and curvature tensor is
	\begin{align*}
		\tilde{g}_{ij} &= \begin{bmatrix} 1+a^2x^2 & -a^2xy \\ -a^2xy & 1+a^2y^2 \end{bmatrix}, \\
		K_{ij} &= \frac{1}{\sqrt{1+a^2(x^2+y^2)}} \begin{bmatrix} -a & 0 \\ 0 & a \end{bmatrix}
	\end{align*}
	We will focus on the saddle point $x=y=0$, where these tensors reduce to $g_{ij} = \text{diag}(1,1)$ and $K_{ij} = \text{diag}(-a,a)$.
	Applying \eqref{eq:modewithcurvature}, we get an expression for a stripe pattern:
	\begin{align*}
		\psi(x,y) &= \psi_0 \cos[q_0(s(x) \cos(\alpha) + s(y) \sin(\alpha))]
	\end{align*}
	where
	\begin{align*}
		s(x) &= \int_0^x \sqrt{g_{xx}} \text{d}x' = \frac{1}{2} x \sqrt{1+a^2x^2} + \frac{1}{2a} \sinh^{-1}(ax)
	\end{align*}
	In this case the finite-thickness energy density $k_{\tilde S}$ as a function of azimuthal angle $\alpha$ reduces to
	\begin{align*}
		k_{\tilde S} &= \frac{h^2}{3 \tilde{A}} \psi_0^2 \cos(4\alpha) + \text{const}.
	\end{align*}
	We see that the minimum energy configuration occurs for $\alpha = \pm \pi/4$.  This fits well with what we find in simulations of the striped phase on a saddle geometry, see Figure \ref{fig:saddle}.

	\subsection{\label{sec:gaussbump} Gaussian bump}
	A Gaussian bump can be parametrized in the following way in Monge gauge:
	\begin{align}
		\tilde{\vec{p}}(r,\phi) &= \begin{bmatrix} r\cos(\phi) & r\sin(\phi) & h_0 \exp[-r^2/(2\sigma^2)]
		\end{bmatrix}^T
	\end{align}
	with $r \geq 0$, $\phi \in [0,2\pi]$.\\
	In general the Gaussian bump has mean curvature as well as Gaussian curvature. However, at the ring given by $r = \sigma$, the Gaussian curvature vanishes and the surface is locally equivalent to a cylinder, with $r$ corresponding to the axis direction and $\phi$ to the azimuthal direction. The model proposed in this paper therefore predicts that there should be a local tendency for to the stripes run \textit{around} the circle as if it was a cylinder.
	
	For a more global view, we must investigate the curvatures of the Gaussian bump in its entirety (not just the $K=0$ circle).
	The two principal curvature directions of the bump are given by $\vec{\hat r}$ and $\boldsymbol{\hat \phi}$. Whenever two non-zero principal curvatures are present, there will correspondingly exist two stripe orientations which corresponds to local minima of the curvature energy. If properly annealed, the stripes should align along the direction of greatest curvature - however this tendency is of course more pronounced when the two curvatures are markedly different.
	
\begin{figure}[htb]
		\includegraphics[width=\columnwidth]{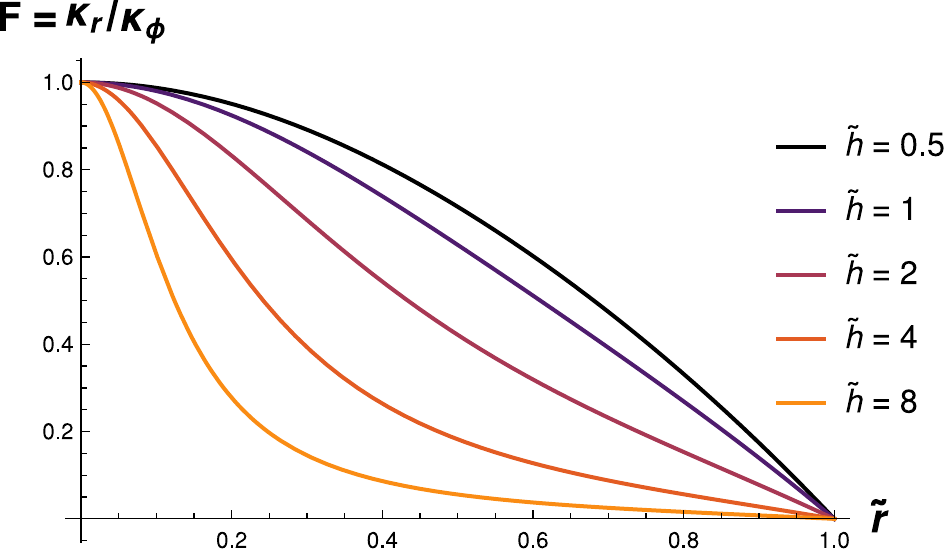}
		\caption{\label{fig:GaussBumpCurvatureRatio} \textbf{Ratio of the principal curvatures on the Gaussian bump}. The azimuthal curvature $K_{\phi\phi}$ is always strongest, regardless of $\tilde r = r/\sigma$ and $\tilde h = h/\sigma$.}
	\end{figure}

	To study the relative strength of the two minimal-energy orientations, we form the ratio of the two principal curvatures $F(\tilde r,\tilde h) := \frac{\kappa_r}{\kappa_\phi}$ where we have defined the two dimensionless quantities $\tilde r = r/\sigma$ and $\tilde h = h_0/\sigma$. One finds:
	\begin{align}
		F(\tilde r, \tilde h) &= \frac{(1+\tilde r)(1-\tilde r)}{1+({\tilde h} {\tilde r})^2 e^{-{\tilde r}^2}}
	\end{align}
	This ratio is plotted in Figure \ref{fig:GaussBumpCurvatureRatio} as a function of $\tilde r$ for several values of the height-to-width ratio $\tilde h$. We see that $F$ drops quickly as a function of $\tilde r$ when $\tilde h$ is large, corresponding to a tall and narrow Gaussian bump. Thus we should see the strongest tendency to orient the stripes azimuthally for narrow and tall Gaussian bumps.
	
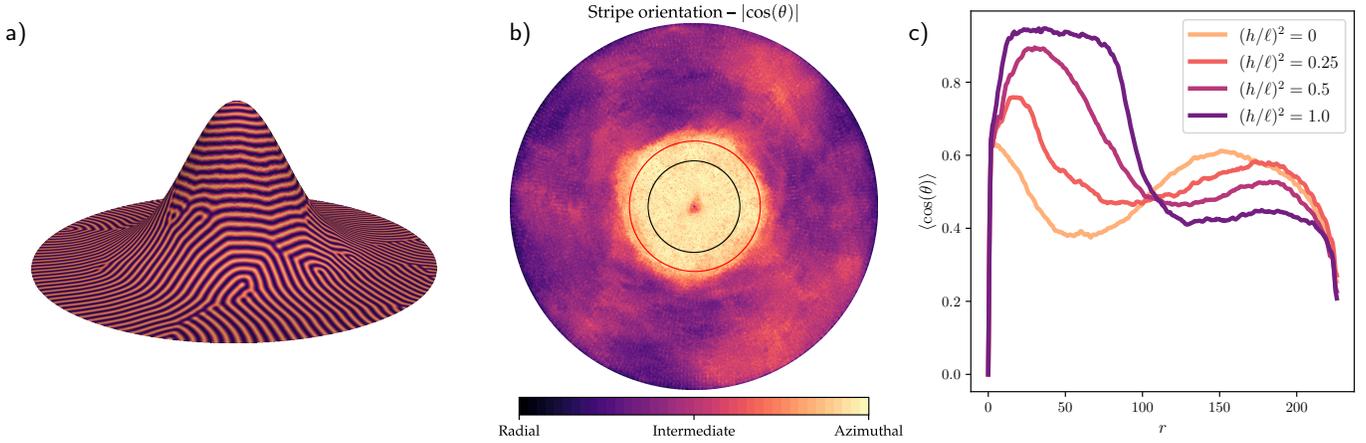
\begin{figure*}[htb]
	\centering
	\begin{tikzpicture}
	    \node at (0,0) {\includegraphics[width=0.6\columnwidth]{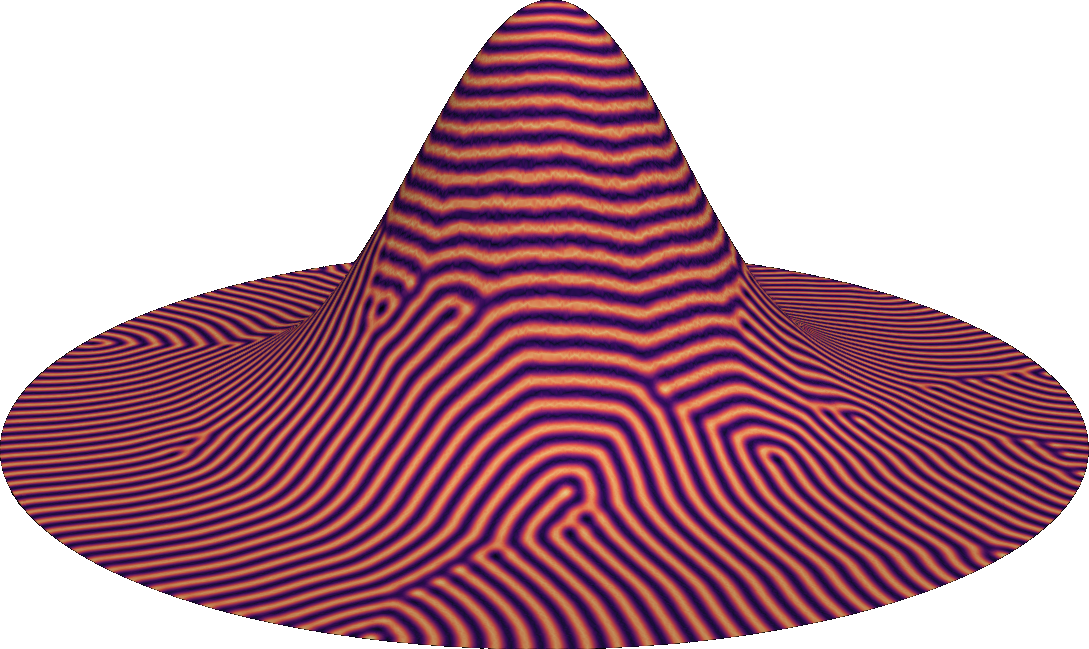}};
	    \node at (6.2,0)
	    {\includegraphics[width=0.6\columnwidth]{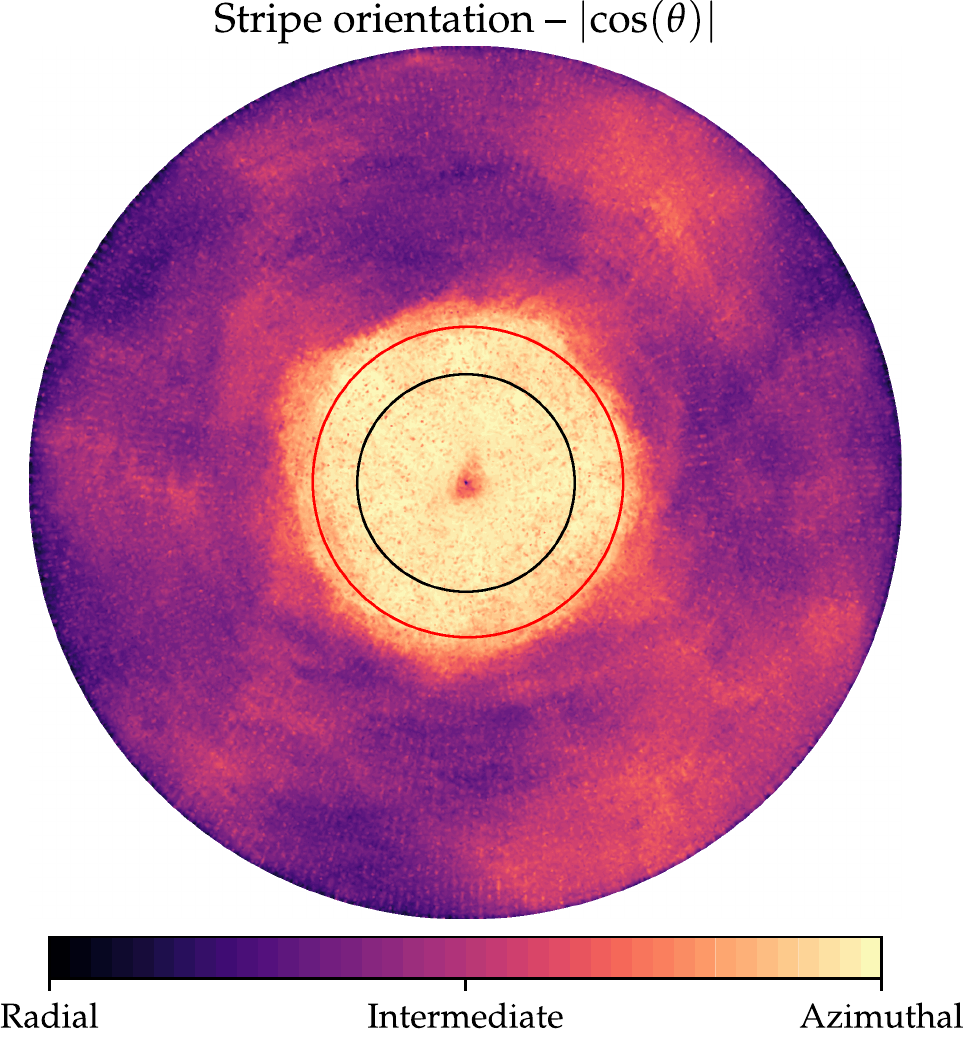}};
	    \node at (12,0)
	    {\includegraphics[width=0.65\columnwidth]{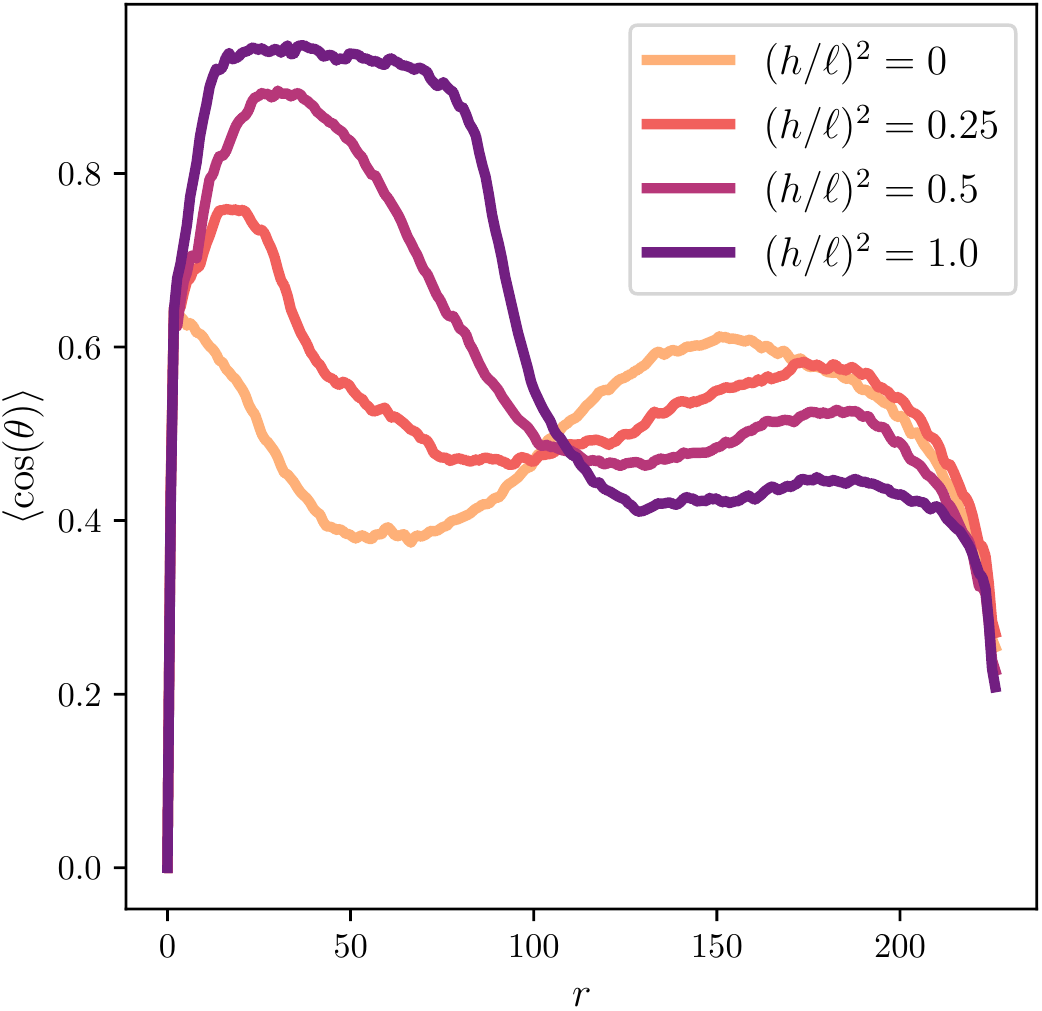}};
	    \node at (-2.9,2.5) {\textbf{\textsf{a)}}};
	    \node at (3.8,2.5) {\textbf{\textsf{b)}}};
	    \node at (9.1,2.5) {\textbf{\textsf{c)}}};
	\end{tikzpicture}
		\caption{\label{fig:GaussBumpPatterns} \textbf{Pattern formation on the Gaussian bump}.
		(a) Stripe pattern formed on Gaussian bump after annealing at $(h/\ell)^2 = 1.0$.
		(b) Average orientation on the Gaussian bump. The color denotes $\left\vert\cos(\theta)\right\vert$ with $\theta = \angle(\hat{\vec{r}},\tilde{\nabla}\psi)$ being the angle between the radial vector and the gradient of $\psi$ along the curved surface. There is a clear tendency for stripes to run azimuthally, \textit{i.e.}\ ``around the bump''. The black circle represents the curve $r = \sigma$ where the Gaussian curvature vanishes. The red circle represents the curve where the two principal curvatures have equal strength, \textit{i.e.} $\left\vert {\kappa_r}/{\kappa_\phi} \right\vert = 1$. For this particular bump, $\tilde h = h_0/\sigma = 3$. This plot is an average over 12 simulations run with $(h/\ell)^2 = 1.0$.
		(c) Average orientation of stripes as a function of radial coordinate. The strength of the orientation effect is clearly controlled by the ratio $(h/\ell)^2$.
		}
	\end{figure*}

	In Fig.\ \ref{fig:GaussBumpPatterns}, the average orientation on the Gaussian bump is shown, with the stripe pattern quite clearly displaying a tendency to run azimuthally (\textit{i.e.} \textit{around} the bump). As with the cylinder, these were obtained by annealing.

	\section{Discussion and conclusions}\label{sec:conclusions}
	Stripe textures of copolymers have frequently been modeled as two-dimensional nematic crystals with a one constant Frank free
	energy.\cite{Nelson2002,Kamien2009,LopezLeon2011} However, due to the use of surface derivatives, this approach does not take the extrinsic geometry into account, and results in all orientations on the cylinder being equivalent. Napoli and Vergori \cite{napoli_extrinsic_2012} considered the influence of extrinsic geometry on a nematic phase on a curved surface, by expanding the three-dimensional Frank elastic energy of nematic crystals to zeroth order in the small parameter $\xi/\ell$ following the technique described in \cite{napoli_surface_2012}. They also considered a cylindrical surface and found that the absolute free energy minimum occurs when the \textit{director field} is aligned with the axis ($\alpha = 0$).
	Thus a similarity between the nematic description and ours arises if the director field is identified with the direction of $\tilde\nabla\psi$.
	A nematic description was also employed by Mbanga \textit{et al.}\cite{Mbanga2012} who studied defects in a nematic phase on a catenoid, and by Segatti \textit{et al.}\cite{Segatti2014} who  investigated  the  behavior  of  the model by Napoli  and  Vergori\cite{napoli_surface_2012,napoli_extrinsic_2012} on a torus. The block copolymer textures considered in this paper are smectic, rather than nematic, but preferred orientations will in many cases turn out to be similar to those obtained in Napoli and Vergori's model \cite{napoli_extrinsic_2012}, due to their model penalizing normal curvature of the director field. However their approach also penalizes any geodesic torsion of the director, something which does not arise in our model. 
	
	Hexemer \cite{Santangelo2007} experimentally studied triblock co-polymer films on an approximately Gaussian bump. In their study, they focused on the orientation at the ring of vanishing Gaussian curvature and found that the stripes tend to be perpendicular to the circle at $r=\sigma$. Our model describes only diblock copolymers, but their result points to the possibility to extend this type of substrate curvature analysis to triblock systems, which have quite different mechanical properties from diblock copolymers.\cite{Matsen1999}
	
	It is clear from our simulations that extrinsic curvature plays a definite role in orienting the pattern. However we should emphasize that our model is purely qualitative, and as such it cannot predict the precise extent of this effect in the laboratory.
	In simulations, the local influence of defects can sometimes overpower the organizing effect of curvature, and thus annealing is usually necessary. Even with annealing, the ordering effect of extrinsic curvature often only becomes significant for relatively large ratios of film thickness to substrate radius of curvature, on the order of $(h/\ell)^2 \sim 0.1 - 1$.
	This should be contrasted with the fact that our perturbative approach is strictly speaking limited to thin films for which the film thickness does not exceed the local radius of curvature. Experiments are often conducted outside this regime -- consider e.g. the experiments of Hexemer \cite{Santangelo2007} in which the film is several layers thick.
	
	To conclude, our study finds that curvature has an important effect on pattern formation in thin film block copolymers. The model that we have developed shows that it is necessary to take the effects of \textit{extrinsic} curvature into consideration -- effects which become apparent due to the finite thickness of the film. Through analysis of three geometries exhibiting distinct signatures of mean vs.\ Gaussian curvature, we conclude that the general tendency is for stripes to align with the direction of maximal curvature. This simple principle provides a straightforward recipe for optimizing the substrate to favour a desired pattern. 
	
	\section*{Conflicts of interest}
	The authors have no conflicts to declare.
	
	\section*{Acknowledgements}
	GL was partly supported by the Research Council of Norway through its Centres of Excellence funding scheme, Project No. 262644.
	
	\appendix
	\section{Numerical method}\label{sec:numerics}
	
	\newcommand{\Sdomain}{\widetilde{\Omega}}
    \newcommand{\domain}{{\Omega}}

    \newcommand{\Sinner}[2]{\left( #1, #2 \right)_{\Sdomain}}
    \newcommand{\inner}[2]{\left( #1, #2 \right)_{\domain}}

    \makeatletter
    \DeclareRobustCommand{\iscircle}{\mathord{\mathpalette\is@circle\relax}}
    \newcommand\is@circle[2]{%
        \begingroup
        \sbox\z@{\raisebox{\depth}{$\m@th#1\bigcirc$}}%
        \sbox\tw@{$#1\square$}%
        \resizebox{!}{\ht\tw@}{\usebox{\z@}}%
        \endgroup
    }   
    \makeatother

    \newcommand{\Sgradvec}{\widetilde{\grad{}}}
    \newcommand{\Sgrad}{\widetilde{\nabla}}
    \newcommand{\Slapl}{\mathop{}\!\widetilde{\bigtriangleup}}
    \newcommand{\Scurvlapl}{\widetilde{\iscircle}}

    \newcommand{\SdA}{\diff \widetilde{S}}
    \newcommand{\dA}{\diff S}

    \newcommand{\sqrtg}{\sqrt{|g|}}
    \newcommand{\KT}{\mathsf{K}}
	
	\newcommand{\pdt}[1]{\partial_t #1}
	
	To solve the relaxation dynamics \eqref{eq:dynamics} with the free energy functional \eqref{eq:f_final}, we have used a numerical framework \cite{surfaise2019} developed by the authors for solving partial differential equations on arbitrary parametrized surfaces.
	The numerical framework is built as a layer on top of the FEniCS/Dolfin framework \cite{logg2012,logg2012b} for solving PDEs using the finite element method. 
	FEniCS interfaces to \textit{e.g.}\ PETSc \cite{petsc2017} for solving large sparse linear systems arising in the finite element method.
	Our numerical framework is not constrained to the equations of motion described here, but the full functionality will be documented and published elsewhere.
	In particular, the framework is accessed using Python and supports arbitrary analytical surface parametrizations. Surface derivatives, which enter in the metric curvature tensor fields, are computed symbolically using SymPy\cite{SymPy}, thereby achieving accuracy only limited by the interpolation onto the unstructured spatial mesh.
	
	In order for the results to be directly reproducible by the reader, the numerical cases presented here are found as example scripts at the GitHub repository \url{https://github.com/gautelinga/surface_pfc/SoftMatter2019}.
	
	\subsection{Functional derivative of the free energy}
	The functional derivative of the free energy $f$ which enters in the equation of motion \eqref{eq:dynamics}, is given by
    \begin{equation}
    \frac{\delta f}{\delta \psi} = \frac{h}{V} \mu ,
    \label{eq:dfdpsi}
    \end{equation}    
    where the \emph{chemical potential} $\mu = \mu_0 + h^2 \mu_2$ can be decomposed into
	\begin{subequations}
    \label{eq:bcp_model}
    \begin{align}
    \mu_0 &= W'(\psi) + 4 \Slapl \psi + 4 \Slapl^2 \psi, \\
    \mu_2 &= \frac{1}{12} K W'(\psi) + Q_1[\psi] + Q_2[\psi] .
    \end{align}
    \end{subequations}
    Here, we have introduced the operators
    \begin{align}
    \Slapl f &= \Sgrad_i \Sgrad^i f 
    & \text{(surface laplacian)} \\
    \Scurvlapl f &= \Sgrad_i ( K^{ij} \Sgrad_j f ) 
    & \text{(curvature laplacian)}
    \end{align}
    and further
    \begin{align}
    Q_1[f] &= \frac{2}{3} \left[ H \Scurvlapl f - K \Slapl f \right], \\
    Q_2[f] &= \frac{1}{3} \left[ 4 \Scurvlapl^2 f - 5 K \Slapl^2 f + 2 H ( \Slapl \Scurvlapl f + \Scurvlapl \Slapl f ) \right],
    \end{align}
    for an arbitrary scalar function $f$.
    Finally, $W'(\psi)$ denotes the derivative of the double well potential $W(\psi) = \tau \psi^2/2 + \psi^4/4$.
    
    \subsection{Time-stepping scheme}
    We discretized the equations of motion \eqref{eq:dynamics} using an implicit approach:
    \begin{equation}
        \frac{\psi^k - \psi^{k-1}}{\Delta t^k} = \tilde M \tilde{\nabla}^2 \mu^k,
    \end{equation}
    where $\psi^k$ approximates $\psi(\tilde{\vec p}, t^k)$, $\mu^k$ approximates $\mu$ at time $t^k$, and the constant mobility $\tilde{M}$ has absorbed the prefactor in \eqref{eq:dfdpsi}, i.e.\ $\tilde M = (h/V) M$ (compare \eqref{eq:dynamics}).
    The time step is given by $\Delta t^k = t^k - t^{k-1}$ and selected adaptively; an initial estimate is based on
    \begin{equation}
        \Delta t^k_* = \frac{c}{\max\{|\nabla \mu^{k-1}|\}}
        \label{eq:dtest}
    \end{equation}
    where $c$ is a heuristically chosen constant.
    An estimate like \eqref{eq:dtest} is fairly standard for phase-field models, cf.\ \cite{campillo-funollet2012}, and it leads to large (small) time steps when the driving forces are small (large).
    If the time step is still too large to achieve convergence within a few iterations, a new time step is chosen as half of the previous estimate, \textit{i.e.}\ $\Delta t_*^k \to \Delta t_*^k/2$, which is repeated until convergence.
    
    The chemical potential is given by
    \begin{equation}
        \mu^k = \mu_0^k + h^2 \mu_2^k,
    \end{equation}
    where
    \begin{subequations}
    \begin{align}
    \mu_0^k &= \overline{W'}(\psi^k, \psi^{k-1}) + 4 \Slapl \psi^k + 4 \Slapl^2 \psi^k, \label{eq:mu0def}\\
    \mu_2^k &= \frac{1}{12} K \overline{W'}(\psi^k, \psi^{k-1}) + Q_1[\psi^k] + Q_2[\psi^k], \label{eq:mu2def}
    \end{align}
    \end{subequations}
    In \eqref{eq:mu0def} and \eqref{eq:mu2def}, the function $\overline{W'}(\psi^k, \psi^{k-1})$ approximates the derivative of the double well potential $W'(\psi)$.
    Herein, we choose the fully implicit nonlinear discretization 
    \begin{equation}
        \overline{W'}(\psi^k, \psi^{k-1}) = W'(\psi^k).
        \label{eq:Wprime}
    \end{equation}
    Apart from the terms involving $W'$, the model is linear. Further, with the choice \eqref{eq:Wprime}, it can be shown that the numerical scheme satisfies a second law of thermodynamics on the discrete level; i.e.\ the discrete free energy (replacing $\psi \to \psi^k$ in \eqref{eq:f_final} decays in time:
    \begin{equation}
        f[\psi^k] \leq f[\psi^{k-1}].
    \end{equation}
    
	\subsection{Variational form}
	The problem is solved using mixed finite elements, all of which belong to the space of piecewise continuous functions.
	In particular, we introduce the auxiliary fields
	\begin{equation}
	    \nu^k = \Slapl \psi^k, \quad \text{and} \quad \hat{\nu}^k = \Scurvlapl \psi^k,
	\end{equation}
	such that our trial functions are given by
	\begin{equation}
	    [\psi^k, \mu^k, \nu^k, \hat{\nu}^k ] \in W = ( H^1(\Omega) )^4.
	\end{equation}
	Further, to save some notation, we define the ``gradient products''
	\begin{align}
    \mathfrak{I}[a, b] &= g^{ij} a_{,i} b_{,j},  \\
    \mathfrak{K}[a, b] &= K^{ij} a_{,i} b_{,j},
    \end{align}
    where $a, b$ are scalar fields.
	
	The variational problem to be solved can now be posed as the following: Given $\psi^{k-1}$, find $[\psi^k, \mu^k, \nu^k, \hat{\nu}^k ] \in W$ such that for all test functions
	$[\chi, \eta, \zeta, \hat{\zeta} ] \in W$, we have
	
	\begin{subequations}
	\begin{align}
    0 &= \int_{\Sdomain} \left[ \frac{\psi^k - \psi^{k-1}}{\Delta t^k} \chi + M \mathfrak{I} [\mu^k, \chi] \right] \, \SdA ,\\
    0 &= \int_{\Sdomain} {\mu^k}{\xi} \, \SdA  - m,\label{eq:varf_2} \\
    0 &= \int_{\Sdomain} \left[ \nu^k\zeta  + \mathfrak{I}[{\psi^k},{\zeta}] \right]\SdA, \\
    0 &= \int_{\Sdomain} \left[ {\hat{\nu}^k}{\hat{\zeta}} + \mathfrak{K}[{\psi^k},{\hat\zeta}] \right]\SdA,
    \end{align}
    \label{eq:fullprob}
    \end{subequations}
    where, in \eqref{eq:varf_2}:
    \begin{subequations}
    \begin{align}
    m &= m_{\textrm{NL}} + m_0 + h^2 m_2.
    \end{align}
    The nonlinear contribution is given by
    \begin{align}
    m_{\textrm{NL}} [\psi^k, \psi^{k-1}, \xi ] 
    &= \int_{\Sdomain} {\left(1 + \frac{h^2}{12}K\right) \overline {W'}(\psi^k, \psi^{k-1})}{\xi} \, \SdA,
    \end{align}
    and the (zeroth and second order in $h$) linear contributions are given by
    \begin{align}
    m_0 [\nu^k, \xi] 
    &= 4 \int_{\Sdomain} \left[ {\nu^k}{\xi} - \mathfrak{I} [{\nu^k},{\xi}] \right] \SdA, \\
   m_2 [\nu^k, \hat{\nu}^k, \xi]
    &=
    \frac{1}{3} 
    \int_{\Sdomain} \Big[ {2} \left[ H \hat{\nu}^k - K \nu^k \right] {\xi} 
    - 4 \mathfrak{K} [{\hat{\nu}^k},{\xi}] + {5 K} \mathfrak{I}[{\nu^k},{\xi}] \nonumber \\ 
     & \qquad \qquad - {2 H} \left( \mathfrak{I}[{\hat{\nu}^k},{\xi}] + \mathfrak{K}[{\nu}^k,{\xi}]\right) \Big] \SdA. 
    \end{align}
    \label{eq:fullprob_extras}
    \end{subequations}
    At each time step $k$, this gives the solution vector $[\psi^k, \mu^k, \nu^k, \hat{\nu}^k]$ wherein $\psi^k$ is used for the next time step $k+1$.
    
    Since the variational problem has a nonlinear contribution from $m_{\rm NL}$, the problem must be linearised and solved in an inner iteration cycle at each time step. 
    In practice, this is automatically handled in FEniCS, which automatically generates the Jacobian of the system based on the symbolic expression for $W'(\psi^k)$ to be used in a Newton method with $\psi^{k-1}$ as an initial guess for $\psi^k$.
    
    The full set of equations \eqref{eq:fullprob} with \eqref{eq:fullprob_extras} was discretized \emph{on the reference domain}, i.e. a linear system was found \textit{i.e.}\ using
    \begin{equation}
        \int_{\Sdomain} (\bullet )\, \SdA = \int_{\domain} (\bullet) \sqrtg \, \dA .
    \end{equation}
    For stability purposes (particularly when $h/\ell$ was large), the best convergence rate was achieved using a direct linear solver.
	
	
	
	\balance
	
	
 \providecommand{\noopsort}[1]{} \providecommand{\singleletter}[1]{#1}%
\newcommand{\acisc}{Adv. Colloid Interface Sci.} \newcommand{\ACMTOMS}{ACM
	Trans. Math. Softw.} \newcommand{\AdvGeophys}{Adv. Geophys.}
\newcommand{\AdvPhys}{Adv. Phys.} \newcommand{\AdvWatRes}{Adv. Water Resour.}
\newcommand{\AIPConfProc}{AIP Conf. Proc.} \newcommand{\AnnPhys}{Ann. Phys.}
\newcommand{\AnnuRevCon}{Annu. Rev. Control} \newcommand{\ApplMechRev}{Appl.
	Mech. Rev.} \newcommand{\ApplPhysLett}{Appl. Phys. Lett.}
\newcommand{\ARFM}{Annu. Rev. Fluid Mech.} \newcommand{\BrazJPhys}{Braz. J.
	Phys.} \newcommand{\BrHeartJ}{Br. Heart J.}
\newcommand{\CommunComputPhys}{Commun. Comput. Phys.}
\newcommand{\CRAcadSci}{C. R. Acad. Sci.} \newcommand{\CRAcadSciIIB}{C. R.
	Acad. Sci. IIB} \newcommand{\CRMecanique}{C. R. M{'e}canique}
\newcommand{\ComputFluids}{Comput. Fluids}
\newcommand{\ComputMethodsApplMechEng}{Comput. Methods Appl. Mech. Eng.}
\newcommand{\cpc}{Comput. Phys. Commun.} \newcommand{\csurfa}{Colloids Surf.
	A} \newcommand{\EarthSciRev}{Earth-Sci. Rev.} \newcommand{\ejam}{Eur. J.
	Appl. Math.} \newcommand{\EnergyFuels}{Energy Fuels}
\newcommand{\EPL}{Europhys. Lett.}
\newcommand{\ESAIMMathModelNumerAnal}{ESAIM Math. Model. Numer. Anal.}
\newcommand{\EurJMB}{Eur. J. Mech. B Fluids} \newcommand{\EurPhysJH}{Eur.
	Phys. J. H} \newcommand{\FaradayDiscuss}{Faraday Discuss.}
\newcommand{\gji}{Geophys. J. Int.} \newcommand{\grl}{Geophys. Res. Lett.}
\newcommand{\ieeetps}{IEEE Trans. Plasma Sci.} \newcommand{\IJGGC}{Int. J.
	Greenh. Gas Con.} \newcommand{\ijhmt}{Int. J. Heat Mass Transf.}
\newcommand{\ijms}{Int. J. Mol. Sci.} \newcommand{\ijmf}{Int. J. Multiph.
	Flow} \newcommand{\ijnmf}{Int. J. Numer. Methods Fluids}
\newcommand{\IJRMMS}{Int. J. Rock Mech. Min. Sci.}
\newcommand{\InterfaceFreeBound}{Interface Free Bound.}
\newcommand{\JAcoustSocAm}{J. Acoust. Soc. Am.} \newcommand{\JAmChemSoc}{J.
	Am. Chem. Soc.} \newcommand{\jap}{J. Appl. Phys.} \newcommand{\jast}{J.
	Adhes. Sci. Technol.} \newcommand{\JChemPhys}{J. Chem. Phys.}
\newcommand{\jcis}{J. Colloid Interface Sci.} \newcommand{\jcompm}{J. Comp.
	Math.} \newcommand{\jcompp}{J. Comput. Phys.} \newcommand{\jfm}{J. Fluid
	Mech.} \newcommand{\jgrse}{J. Geophys. Res. Solid Earth}
\newcommand{\JHydrol}{J. Hydrol.} \newcommand{\jmems}{J. Microelectromech.
	Syst.} \newcommand{\jpc}{J. Phys. Chem.} \newcommand{\JPhysChemB}{J. Phys.
	Chem. B} \newcommand{\jpcm}{J. Phys. Condens. Matter}
\newcommand{\JPetSciEng}{J. Pet. Sci. Eng.} \newcommand{\JPSciEng}{J. Pet.
	Sci. Eng.} \newcommand{\JStatPhys}{J. Stat. Phys.} \newcommand{\labchip}{Lab
	Chip} \newcommand{\MarPetGeol}{Mar. Pet. Geol.} \newcommand{\MathComp}{Math.
	Comp.} \newcommand{\MathProcCambPhilosSoc}{Math. Proc. Camb. Philos. Soc.}
\newcommand{\mmmas}{Math. Models Methods Appl. Sci.}
\newcommand{\natgeo}{Nat. Geosci.} \newcommand{\natphys}{Nat. Phys.}
\newcommand{\nedes}{Nuclear Engineering and Design}
\newcommand{\NumerMath}{Numer. Math.} \newcommand{\pageoph}{Pure. Appl.
	Geophys.} \newcommand{\PhilosMag}{Philos. Mag.}
\newcommand{\physicaa}{Physica A} \newcommand{\physicad}{Physica D}
\newcommand{\physfluids}{Phys. Fluids} \newcommand{\physfluidsa}{Phys. Fluids
	A} \newcommand{\PhysRep}{Phys. Rep.} \newcommand{\PhysRev}{Phys. Rev.}
\newcommand{\physreve}{Phys. Rev. E} \newcommand{\PhysRevFluids}{Phys. Rev.
	Fluids} \newcommand{\physrevlett}{Phys. Rev. Lett.} \newcommand{\PhysZ}{Phyz.
	Z.} \newcommand{\PNAS}{Proc. Natl. Acad. Sci.} \newcommand{\prsa}{Proc. Royal
	Soc. A} \newcommand{\ptrsl}{Philos. Trans. R. Soc.}
\newcommand{\ptrsla}{Philos. Trans. R. Soc. A} \newcommand{\RepProgPhys}{Rep.
	Prog. Phys.} \newcommand{\RevGeophys}{Rev. Geophys.}
\newcommand{\revmodphys}{Rev. Mod. Phys.} \newcommand{\SciRep}{Sci. Rep.}
\newcommand{\SIAP}{SIAM J. Appl. Math.} \newcommand{\SINUM}{SIAM J. Numer.
	Anal.} \newcommand{\SISC}{SIAM J. Sci. Comput.} \newcommand{\SJSSC}{SIAM J.
	Sci. and Stat. Comput.} \newcommand{\transpporousmedia}{Transp. Porous Media}
\newcommand{\waterresres}{Water Resour. Res.} \newcommand{\ZPhysB}{Z. Phys.
	B} \newcommand{\ZPhysChem}{Z. Phys. Chem.}
\providecommand*{\mcitethebibliography}{\thebibliography}
\csname @ifundefined\endcsname{endmcitethebibliography}
{\let\endmcitethebibliography\endthebibliography}{}

\end{document}